\documentclass{aa}  
\usepackage{hyperref}
\usepackage{stfloats}
\usepackage{lscape}
\usepackage{natbib}
\usepackage{array} 
\usepackage[utf8]{inputenc}
\usepackage{amsmath,amssymb}
\usepackage{xcolor}
\usepackage{float}
\usepackage{graphicx}
\usepackage{txfonts}
\defcitealias{jimenez2024kratos}{JA24}

\begin{document} 

   \title{Tidally-induced radial migration waves in LMC-like galaxies}

   \author{D. Hebrail
          \inst{1,2}
          \and
          Ó. Jiménez-Arranz
          \inst{2}
          \and
          S. Roca-Fàbrega
          \inst{2}$^,$ \inst{3}
          }

   \institute{Université de Toulouse, 118 route de Narbonne, 31062 Toulouse Cedex 9, France
   \\
              \email{dylan.hebrail@fysik.lu.se}
         \and
             Lund Observatory, Division of Astrophysics, Lund University, Box 43, SE-221 00 Lund, Sweden
    \and
             Departamento de Física de la Tierra y Astrofísica, Fac. de C.C. Físicas, Universidad Complutense de Madrid, E-28040 Madrid, Spain\\
             }

   \date{Received <> / Accepted <>}

 
  \abstract
  {Stellar radial migration has predominantly been examined in isolated disc galaxies where non-axisymmetric structures drive the process. By contrast, while tidal interactions are known for having an influence, their contribution remains comparatively under explored. The LMC, the nearest disc galaxy to the Milky Way (MW) and currently interacting with the SMC, provides a unique laboratory to investigate this interplay.}
   {We aim to quantify the impact of tidal interactions on radial migration and metallicity distribution in high-resolution simulations of LMC-like disc galaxies.}
  {We leverage a subsample of KRATOS, a suite of 28 pure $N$-body simulations of the LMC-SMC-MW system. Specifically, we use 6 simulations of both isolated and interacting LMC-like galaxies, exploring different values of the Toomre stellar parameter $Q$. These simulations allow to map the evolution of the stars' guiding radii $R_g(t)$ and compute radial migration fluxes in interacting systems and compare with their isolated counterparts, allowing to quantify the link between tidal interactions, radial migration, non-axisymmetric patterns, disc internal stability, and radial metallicity distribution.}
   {We present tidally-triggered wave-like radial migration fluxes reaching up to $\sim40\%$ of disc stellar mass per Gyr. This wave-like migration appears during the satellite's pericentre passages, almost independently of $Q$ and induces a metallicity drop of $\sim$3-5\% of the isolated galaxy's maximum metallicity in the inner disc. Additionally, in the isolated simulations, the extent of variation in the bar's resonance region coincides with the mixing zones in the metallicity distribution.}
   {We propose a novel description of a wave-like radial migration flux as a dynamical response of a galaxy undergoing tidal interactions and sketch its impact on the galaxy's metallicity distribution.}


   \maketitle
%

\section{Introduction}
Disc galaxies, characterized by their flattened stellar distributions and rotationally supported dynamics, represent more than half of the total local galactic population \citep[e.g.,][]{zoo1,zoo2}. They consist of stars, gas, dust, and dark matter, and often display non-axisymmetric structures as spiral arms and bars \citep[e.g.,][]{DEVAUCOULEURS1972163, zoo09}. It has long been known that galaxies were likely to merge together during the lifetime of our universe \citep{tremaine1981galaxy}, further compounding the natural complexity of these large scale systems. Therefore, to gain insight into galactic dynamics, the study of galactic interactions is essential. 

Because of their large spatial extent, galaxies experience non-uniform gravitational forces during interactions, which drive both their relative motion and their internal reshaping. The effect induced by the gradient of the gravitational force is called a tidal interaction and, as presented in the seminal article by \citet{1972ApJ...178..623T}, can significantly transform galaxies. Further research allowed to demonstrate the ability of tidal interactions to induce star formation bursts in galaxies \citep[e.g.,][]{lotz2008galaxy}, drastically change their morphology \citep[e.g.,][]{white1978simulations}, heat and thicken their disc, strongly weaken radial metallicity gradients \citep[e.g.,][]{BustamanteMergerMetal}, displace their centre of mass \citep[i.e. creating a ``reflex motion'', e.g.,][]{Erkalreflex19,Petersenreflex21,BrooksReflex2025,Yaaqibreflex25}, impact the morphology and kinematics of nearby stellar streams \citep[e.g.,][]{erkal19streams, Kosopov23streams, brooks24streams, Jimenez25}, and create new elliptical galaxies \citep[e.g.,][]{BournaudMergerEllip}, among others. Thus, understanding the dynamical response of galaxies to tidal interactions constitutes a major objective in contemporary astrophysics.
 
Tidal interactions are not the only mechanism able to perturb a galactic disc. Indeed, the interplay between non-axisymmetric patterns and the broad variety of coexisting stellar orbits allows to redistribute angular momentum in disc galaxies. Thus, some stars can undergo significant and durable radial excursions: this mechanism is known as radial migration \citep{sellwood2002radial}. The key quantity to quantify the radial migration of a star orbiting within a disc galaxy is its guiding radius $R_g(t)$: the equivalent radius of a circular orbit of a star with the same angular momentum. 
In practice, the variations of the guiding radii of stars in the galactic disc is called churning \citep[as introduced in][]{schonrich2009chemical}, and directly implies a redistribution of the stellar population in the galaxy. This churning was initially described as caused by transient spiral arms formation \citep[see][for a review on the topic]{sellwood2014secular}, but is now also linked to bar instabilities, and resonances with rotating patterns in disc galaxies \citep[e.g.,][]{minchev2011radial, halle2015quantifying, nagy2025comparing, marques2025bar}. Regarding examples of its consequences: radial migration can significantly weaken the radial metallicity gradient of galaxies \citep[e.g.,][]{10.1093/mnras/stt1667, gradflat15} and increase the metallicity dispersion in the disc \citep[e.g.,][]{sellwood2002radial, 10.1093/mnras/stv016}.

In this context, radial migration has been primarily examined in simulations of isolated disc galaxies \citep[e.g.,][]{halle2015quantifying, vera2016imprint, halle2018thickdisc, mikkola2020radial, iles2024impact}, and in a lesser extent, with satellite galaxies \citep[e.g.][]{quillen2009radial,Bird2012}. More recently radial migration is investigated within MW-Sagittarius \citep[e.g.,][]{Carr2022} and cosmological \citep[e.g.,][]{Lu24,ImpactRM.I,ImpactRM.II} simulations, where the influence of tidal interactions on stellar churning is explicitly considered. These latter works show that tidally induced migration can significantly modify the inferred star formation histories of galaxies, as well as other key observational properties.
However, while cosmological simulations have provided insight into radial migration, their limited resolution and complex environments make systematic investigations difficult. Thus, the impact of interactions on radial migration in high-resolution simulations remains insufficiently addressed, forming the basis for our analysis.

On the other hand, we see that our neighbourhood is no stranger to tidal interactions, with the Milky Way (MW) history including numerous mergers and satellite galaxies disruptions \citep[e.g.,][]{10.1093/mnras/278.3.727, antoja2018dynamically, merrow2024did,Hunt2025review}. Among these merged satellites, notable examples are the Sagittarius dwarf spheroidal galaxy \citep[Sgr dSph, see][]{sagittarius} and the Gaia Enceladus/Sausage \citep[GES, see][]{HelmiGES, BelokurovGES}. In addition to this merger history, several satellite galaxies populate the surroundings of our Galaxy.
Among these, the LMC — the closest disc galaxy to the MW, located approximately 50 kpc from the Sun \citep{graczyk2013araucaria, pietrzynski2019distance} and currently interacting with the SMC — represents a particularly relevant target for our study.

Because of this proximity, the LMC is an uniquely suited laboratory for diagnosing radial migration, especially under the influence of tidal interactions. It is a warped barred Magellanic Spiral \citep{DEVAUCOULEURS1972163}, featuring a single prominent spiral arm connected to its slightly off-centred bar \citep{el2019vmcLMCmorph, luri2021gaia, jimenez2023kinematic,  jimenezarranz2025verticalstructurekinematicslmc, Rathore25} that has been recently measured as non-rotating \citep{jimenez2024bar}. 
The LMC is undergoing tidal interactions with the SMC, providing a natural explanation for the LMC's non-rotating bar \citep[see][]{2025jimenezbarstop}, and the MW \citep[e.g.,][]{LMCwake,vasiliev2023effect}.
Owing to its favourable position in the sky and proximity, the LMC has been the target of numerous observational surveys \citep[e.g.,][]{MCstudy1, MCstudy2, 4most, MCstudy3}. With $Gaia$ resolving around $\sim$15–20 million of its stars \citep{luri2021gaia, jimenez2023kinematic, jimenez2023b,Perryman25gaia}, the system is not only exceptionally well characterized observationally but has also been the subject of extensive numerical modelling \citep[e.g.,][]{Besla12, LMCwake,Brooks25, Sheng25LMCperturbhaloMW}. In particular, in this work we use the KRATOS suite of simulations \citepalias{jimenez2024kratos} to characterize how tidal interactions drive radial migration in LMC-like systems.

The KRATOS suite comprises 28 pure $N$-body simulations of the LMC–SMC–MW system, which naturally reproduce several observed features of the LMC, including the offset of its central non-rotating bar \citep{2025jimenezbarstop}, the warped structure of its disc \citep{jimenezarranz2025verticalstructurekinematicslmc}, and its kinematic asymmetries \citep{MarieAsymmetries25}. The diverse initial conditions across the simulations yield a variety of stellar bar and spiral arm morphologies in the LMC-like systems, enabling an extension of the analysis to the broader subclass of Magellanic spiral galaxies. Furthermore, KRATOS has demonstrated that tidal interactions in LMC analogues can trigger a wide array of mechanisms driving radial migration \citepalias[see the main work of KRATOS:][]{jimenez2024kratos}, including connection between bar and spiral arms \citep[e.g.,][]{marques2025bar}, and changes in the bar pattern speed during mergers \citep[e.g.,][]{zhang2025radial, 2025jimenezbarstop}. Consequently, these simulations are expected to host a rich diversity of churning modes, making them particularly well suited for quantifying the impact of tidal interactions on radial migration in disc galaxies. Building on KRATOS, we aim to provide a novel characterization of how tidal interactions influence the radial redistribution of both stars and metallicities in disc galaxies.

To this end, we organised the paper as follows. In Sect.~\ref{sec:kratos}, we describe the KRATOS simulations as a basis for this work. In Sect.~\ref{sec:methods}, we present the methodology and assumptions used to quantify radial migration from the KRATOS output data. In Sect.~\ref{sec:results}, we present our results including a tidally-induced regime of oscillating churning fluxes in LMC-like galaxies. In Sect.~\ref{sec:discussion}, we discuss our results, the limitations of our assumptions and the future prospects of this field, before concluding this paper in Sect.~\ref{sec:conclusions}.

\section{The KRATOS suite of simulations}
\label{sec:kratos}

KRATOS \citepalias[originally presented in][]{jimenez2024kratos} is a high-resolution suite of 28 simulations aiming to study the LMC-SMC-MW interacting system. These simulations are pure $N$-body simulations, namely, without hydrodynamics and stellar evolution or formation. This suite of 28 simulations allows to compare the evolution of an isolated LMC-like galaxy with its counterpart, interacting with an SMC-mass and a MW-mass galaxy. In KRATOS, the explored parameters are the mass of the baryonic matter or dark matter content of the different systems and the stellar Toomre parameter Q of the LMC-like system \citepalias[see
Table 2 of][for a summary]{jimenez2024kratos}. In this work we follow the notation introduced in \citetalias{jimenez2024kratos}, where the LMC-like galaxy is denoted as $G_{\rm LMC}$, and the SMC- and MW-mass systems as $G_{\rm SMC}$ and $G_{\rm MW}$, respectively.

The simulations feature the stellar mass resolution of $4\times10^3M_\odot$, along with a spatial and temporal resolutions of 10 pc and 5,000 yr. In the KRATOS suite, none of the galaxies is modelled as an analytical potential and they are simulated for 4.68 Gyr within a $2.85^3$ Mpc$^3$ box with periodic boundary conditions, using the adaptive refinement mesh Eulerian code ART presented in \citet{kravtsov1997adaptive}. For this work, we focus on a subsample of 6 simulations of KRATOS  (see Table~\ref{table:KRATOS}). They were re-ran so that each simulation provides more than $2,000$ snapshots\footnote{The original KRATOS simulations included 61 snapshots per simulation that are available online. The higher temporal cadence version of the simulations listed in Table~\ref{table:KRATOS} can be provided upon reasonable request.} of the evolution of the systems, with a time separation lower than 2.6 Myr. The present time $t=0$ is defined 4.0 Gyr after the initial conditions, and was determined to match
the number of $G_{\rm LMC}$–$G_{\rm SMC}$ pericentre passages and the morphology of the $G_{\rm LMC}$ disc \citepalias[see Sect. 3 of][for more details]{jimenez2024kratos}. For reference, we mention that the time $t=0$ was re-evaluated in Sect. 4.3 of \citet{MarieAsymmetries25} to match the kinematic asymmetries of the disc, but we use only the original definition. The subsample of the KRATOS simulations used in this work may include the following systems: 

\begin{table}
\caption{Subsample of the KRATOS simulations used in this work.}
\label{table:KRATOS}      
\centering    
\begin{tabular}{>{\centering\arraybackslash}p{1.5cm} >{\centering\arraybackslash}p{2cm} p{3cm}}      
\hline\hline                 
Simulation name & Toomre parameter $Q$ & Configuration \\    
\hline                        
   K1 & 1.2 & $G_{\rm LMC}$\\      
   K3\textsubscript{i} & 1.2 & $G_{\rm LMC}+G_{\rm SMC}+G_{\rm MW}$ \\
   K4 & 1.0 & $G_{\rm LMC}$ \\
   K6\textsubscript{i} & 1.0 & $G_{\rm LMC}+G_{\rm SMC}+G_{\rm MW}$ \\
   K7 & 1.5 & $G_{\rm LMC}$\\ 
   K9\textsubscript{i} & 1.5 & $G_{\rm LMC}+G_{\rm SMC}+G_{\rm MW}$\\
\hline                                   
\end{tabular}
\end{table}

 \begin{itemize}
     \item $G_{\rm LMC}$: a LMC-like (i.e. similar in both mass and shape to the LMC) galaxy with a stellar mass of $5.0\times10^9M_\odot$ and a dark matter mass of $1.8\times10^{11}M_\odot$. It features $1.2\times10^6$ stellar particles with a stellar mass resolution of $4\times10^3M_\odot$. $G_{\rm LMC}$ presents an exponential stellar disc profile with a scale height and length of, respectively, 0.20 kpc and 2.85 kpc truncated at 11.5 kpc from the centre, embedded in a dark matter Navarro-Frenk-White (NFW) halo \citep{NFW} that evolves during the simulations. 
     \item $G_{\rm SMC}$: a SMC-mass (i.e. only aiming to reproduce the mass of the SMC) galaxy with a stellar mass of $2.6\times10^8M_\odot$ and a dark matter mass of $1.9\times10^{10}M_\odot$ with the same stellar mass resolution of $4\times10^3M_\odot$. Both stars and dark matter follow a NFW profile \citep[presented in Sect. 2.2 of][]{jimenez2024kratos}. 
     \item $G_{\rm MW}$: a MW-mass galaxy without stellar particles, with a total mass of $10^{12}M_\odot$ distributed in a NFW dark matter halo profile. The mass resolution of its dark matter particles is $5\times10^5M_\odot$. 
 \end{itemize}

The KRATOS suite includes several groups of simulations featuring an isolated and an interacting group of simulations. The isolated group serves as a control group for this work, including only $G_{\rm LMC}$ and allowing to see its secular evolution free from tidal interactions. On the other hand, the interacting group embeds $G_{\rm LMC}$ within a realistic environment, interacting with the two other systems $G_{\rm SMC}$ and $G_{\rm MW}$. We mention that the KRATOS suite includes another group of simulations featuring $G_{\rm LMC}$ and $G_{\rm SMC}$ without $G_{\rm MW}$. We do not use this group for this work and instead, we focus on the control isolated group and the most realistic interacting group of simulations including all three systems. In the interacting simulations, $G_{\rm SMC}$ follows an orbit qualitatively matching the interaction history between the SMC and the LMC \citepalias[see Sect. 3 of][for a detailed description]{jimenez2024kratos}, while $G_{\rm MW}$ tidally interacts with $G_{\rm LMC}$ with a longer range.

In both the isolated and interacting simulation groups, $G_{\rm LMC}$ is the focus of this study, allowing a comparison of its stellar radial migration across the two scenarios. To that end, we focus on the subsample of the KRATOS simulations presented in Table~\ref{table:KRATOS}. It includes three isolated and three interacting simulations, each with a Toomre stability parameter of $Q=1.0$, $Q=1.2$, or $Q=1.5$ (K4, K1, K7 for the isolated simulations, and K6\textsubscript{i}, K3\textsubscript{i}, K9\textsubscript{i} for the interacting simulations\footnote{To identify which simulations correspond to the interacting scenario, we added an ``i'' as a subscript to the relevant model names.}, respectively for the different $Q$ values). In isolated simulations, the least stable discs (K1 and K4, for instance) are more prone to create non-axisymmetric patterns, and the most stable isolated disc (K7) is not expected to create strong non-axisymmetric patterns. Varying $Q$ therefore allows to study the impact of non-axisymmetric patterns and stability on radial migration. In summary, the only varying parameters in our work are $Q$ and the fact that the simulations are interacting or isolated. Thus, all masses and orbits (in interacting simulations) are identical, allowing a systematic study.

\section{Methods}
\label{sec:methods}
With the aim of quantifying the amount of stars undergoing radial migration, we first present the method to define a $G_{\rm LMC}$ in-plane reference frame and infer their guiding radius $R_g(t)$ in Sect.~\ref{sub:refrg}. Then, we present in Sect.~\ref{sub:methodMaps} the methods used to map the non-axisymmetric pattern strengths, the radial migration fluxes (or churning fluxes), and their impact on the metallicity distribution in the disc.

\subsection{Defining a $G_{\rm LMC}$ in-plane reference frame and computing the guiding radius $R_g(t)$}
\label{sub:refrg}
In order to quantify properly dynamical effects, it is first required to define a frame of reference, indeed, especially in the simulations including interactions between  $G_{\rm LMC}+G_{\rm SMC}+G_{\rm MW}$, $G_{\rm LMC}$'s galactic plane gets tilted and sometimes slightly warped due to the interactions \citep[also shown in][]{jimenezarranz2025verticalstructurekinematicslmc}. Since warping is a direct consequence of galactic encounters, its effects cannot be disentangled from those of tidal interactions in this work. Instead, they are treated as a degenerate problem and considered together. However, we correct for tilting, keeping the original shape of $G_{\rm LMC}$ and its vertical structure. 

For each simulated snapshot, we fit a plane through the stellar particles of $G_{\rm LMC}$'s disc, and define the $z$-axis normal to this plane. Then, we define an orthonormal reference frame with the $x$- and $y$-axes within the fitted plane, centred on $G_{\rm LMC}$'s centre of stellar mass \citep[as in][]{MarieAsymmetries25}. This method is slightly affected by the disc's tangled warp, but stands as a good compromise to study $G_{\rm LMC}$ without artificially flattening it, therefore impacting its dynamics. In parallel, we checked the consistency of the method with a plane defined perpendicular to the total angular momentum \Vec{L} of $G_{\rm LMC}$'s stellar particles \citep[as used in][]{MarieAsymmetries25} and found less than 3° of difference between the two planes' inclinations, at all times in all simulations. In the end, chose to keep the plane fitting method because it appeared more stable especially during the interactions.

\begin{figure}
    \centering
    \includegraphics[width=0.88\linewidth]{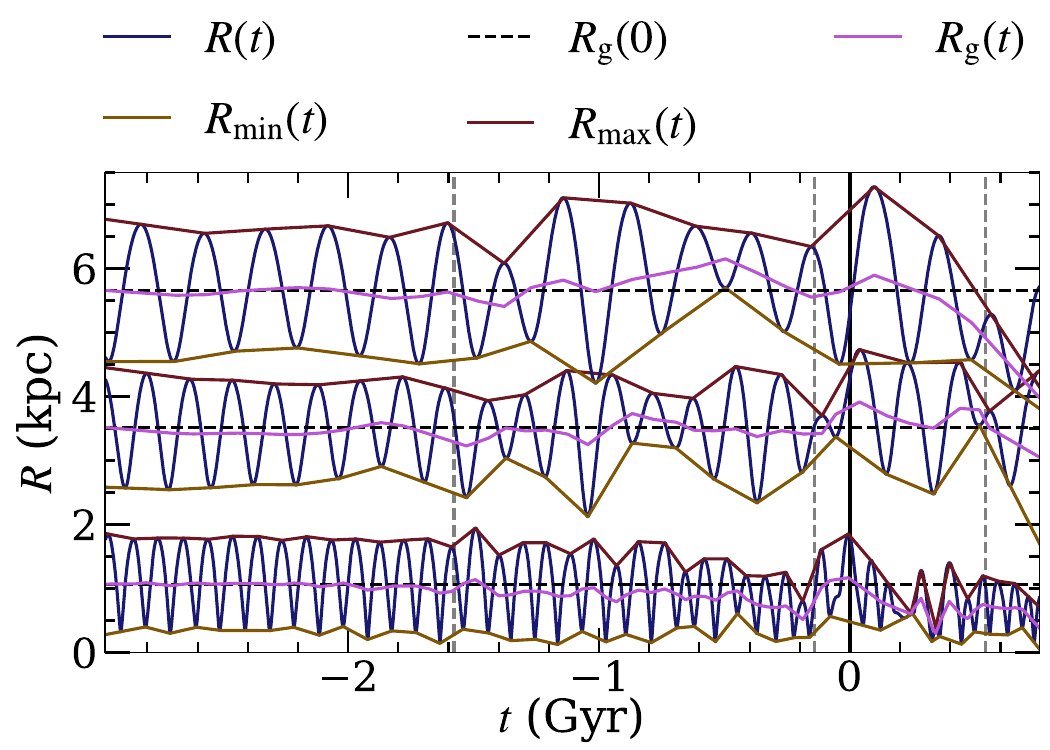}
    \vskip 0.3cm
    \includegraphics[width=0.85\linewidth]{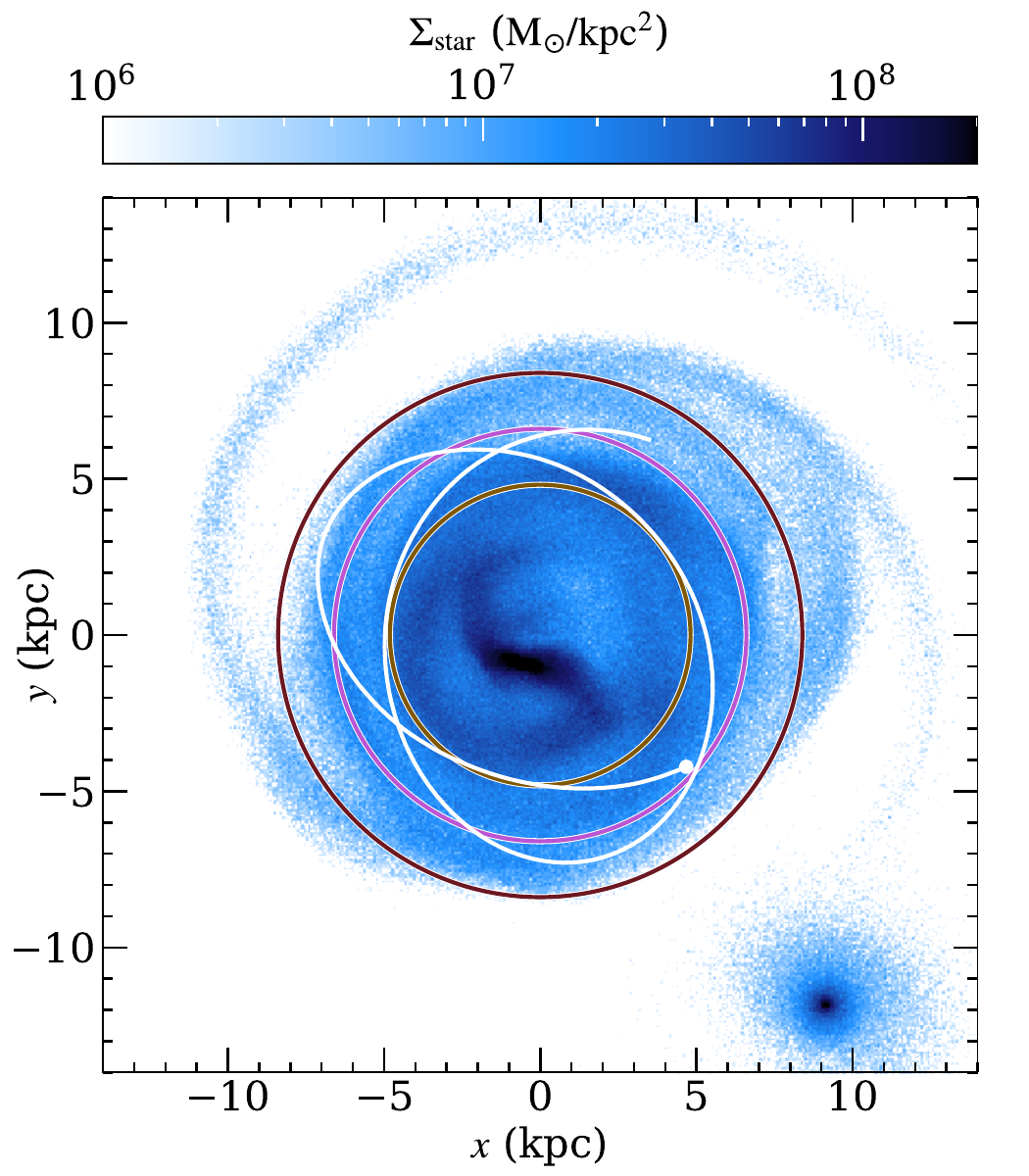}
    \caption{
    Top: Visualisation of the method described in \cite{halle2015quantifying} to infer the $R_g (t)$ of the particles. Radial evolution of three random particles in the K9\textsubscript{i} simulation with respect to time. $R_g(t)$ (pink) is the mean of $R_{\rm max}(t)$ (dark brown) and $R_{\rm min}(t)$ (light brown), and its value at the start of the simulation is $R_g(t_i)$ (black horizontal dashed line). The times of the pericentre passages of $G_{\rm SMC}$ are shown in grey vertical dashed lines, and the evolution of the galactocentric radius of the particles $R(t)$ is the dark blue curve. The black vertical line represents the $t=0$ snapshot that is displayed on the bottom panel.  
    Bottom: Visualisation of the method inferring the guiding radius $R_g(t)$ of a particle at $t=0$ (same legend as above, with the previous orbit of the particle in white). The shown particle is the particle with the highest radius of the top panel. $G_{SMC}$ is shown on the lower right hand corner but its particles are not tracked in this work. An animated visualisation is available at \url{https://youtu.be/SvtrVxrCpwI}.}
    \label{fig:rad_ev_Rg}
\end{figure}

The next step consists in reliably estimating the guiding radius $R_g(t)$ of a star particle in the disc. To do so, we leverage the method presented in \citet{halle2015quantifying}, and used in more recent works such as \citet{khoperskov2020escapees} and \citet{nagy2025comparing}. This method consists in following the evolution of the galactocentric radius of a particle $R(t)$ (see the dark blue curves for three random particles in the disc in Fig.~\ref{fig:rad_ev_Rg}, top panel) and record its local maxima and minima. Then, interpolate two curves through the extrema $R_{\rm min}(t)$ and $R_{\rm max}(t)$ (light and dark brown curves, respectively) and estimate $R_g(t)$ as the mean between these curves  (pink curves). Because of the simplicity of the method, the $R_g(t)$ estimate remains accurate and stable versus most sudden changes in the gravitational potential, galaxy morphology and tidal interactions with another galaxy. The only requirement is a sufficiently sharp time resolution to properly resolve all epicyclic oscillations of the particles. This condition is fulfilled by the KRATOS suite with a time separation lower than 2.6 Myr between all snapshots (see also the oscillations of the particle closest to the centre in Fig.~\ref{fig:rad_ev_Rg}, top panel).

However, this method comes with a two main drawbacks. Indeed, a first unexpected behaviour of the method can occur when $R(t)$ takes an irregular shape (i.e. non-pseudo-oscillating). These orbits occur mostly during strong tidal interactions, but can also punctually be seen in low $Q$ simulations (i.e. K4, K6\textsubscript{i}, and, in a lesser extent, K1, K3\textsubscript{i}) due to their unstable nature. In that case, the particles can reach local extrema at positions that do not match their epicyclic oscillations. In most orbits, these extrema do not stand as problems to estimate $R_g(t)$ because of the stability of the method. For instance, see the orbit of the middle particle of Fig.~\ref{fig:rad_ev_Rg}, top panel: the orbit is irregular during the two last pericentre passages of $G_{\rm SMC}$ (grey vertical dashed lines) but the pink curve still gives a good estimate of the centre of the orbit. However, the $R_g(t)$ estimate of the particle closer to the galactic centre in the same figure presents unrealistic oscillations between the two last pericentre passages. This effect is over-represented in the very centre of $G_{\rm LMC}$'s galactic disc. For this reason and because some slight miscentring of the reference frame can occur, we do not study the radial migration in the region within 1 kpc from $G_{\rm LMC}$'s centre.

The second main drawback of the method is its failure to reliably estimate $R_{\rm min}(t)$, $R_{\rm max}(t)$ and $R_g(t)$ in short time spans close to the time boundaries of the simulations. Indeed, the interpolation of $R_{\rm min}(t)$ and $R_{\rm max}(t)$ has to extrapolate the existing curve when the next (or previous) extrema is located after (or before) the time boundaries, subsequently affecting $R_g(t)$ in these time spans. Because of this, we avoid drawing strong conclusions on radial migration occurring during the last $\sim200$ Myr of the simulations. The first 1.6 Gyr after the initial conditions are also excluded because some galaxies are still in an unrealistic transient state at the very start of the simulations.

Despite these caveats, this method illustrated itself as the best compromise to study radial migration in $G_{\rm LMC}$. We therefore use it to map its galactic disc by estimating the $R_g(t)$ of a large sample of star particles of the disc. Because the KRATOS simulations feature $1.2\times10^6$ stellar particles in the $G_{\rm LMC}$ evolving in more than $2,000$ snapshots, we selected a sample of $1\%$ of all the star particles. We checked the consistency of our results by trying to select different stars and sample sizes, and concluded that $1.2\times10^4$ stellar particles is a sample size large enough to reliably map $G_{\rm LMC}$'s disc.

\subsection{Maps of $G_{\rm LMC}$}
\label{sub:methodMaps}
Since its first description in \citet{sellwood2002radial}, radial migration has been predominantly attributed to non-axisymmetric patterns and recognized as a driver of radial mixing in the disc \citep[e.g.,][]{Minchev_2010, zhang2025radial}. In order to enlighten this interplay, we present three types of maps allowing to survey the interior of $G_{\rm LMC}$'s disc.

First, we map the non-axisymmetric pattern strength of $G_{\rm LMC}$. To that end, we calculate the azimuthal $m=2$ density Fourier mode $A_2$ at each time by integrating it for 30 concentric 0.4 kpc wide rings from the centre of $G_{\rm LMC}$ to $R=12$ kpc. We normalise by the $m=0$ density mode $A_0$, being the total integrated mass for those ring. Therefore, $A_2/A_0$ is the fraction of the bisymmetric component over the total mass of each bin.

Second, we map net churning fluxes within $G_{\rm LMC}$ in order to quantify radial migration in the disc \citep[see][for a slightly different implementation of this method]{halle2015quantifying}. A churning flux (also named radial migration flux) at a radius $R$ is the fraction of disc's stellar mass whose $R_g(t)$ crosses $R$ between two snapshots, divided by the time separation between these snapshots. The net outward flux is counted positive, and inward flux is counted negative. We present churning fluxes with respect to both $t$ and $R$ by calculating them between every snapshot of the simulations. The high time cadency allows to fully resolve the global tendencies of radial migration during the interactions. On the other hand, the radial mixing in  $G_{\rm LMC}$ is less well described by the net churning fluxes due to the cancelling effect between the outward and inward migration. This effect, however, is better traced by the metallicity evolution of the disc, whose method of study is described hereafter.

Third, we map the metallicity distribution within $G_{\rm LMC}$'s disc. Radial migration has long been described as a driver of radial metallicity gradient weakening \citep[e.g.,][]{schonrich2009chemical, 10.1093/mnras/stt1667, 10.1093/mnras/stv016}. We aim to verify that statement and extend the study to interacting simulations by quantifying the impact of radial migration on the metallicity distribution in the KRATOS suite. Because the KRATOS simulations do not originally consider any form of metallicity, star formation or stellar evolution, some assumptions are necessary.
 
Firstly, we model stellar particles with a fixed metallicity throughout the simulations. This is justified by the fact that we only aim to quantify the impact of churning on a disc's metallicity profile. Furthermore, we study migration in a total timespan of $3.68$ Gyr, which is short enough to consider radial migration as the dominating metallicity-impacting mechanism. Secondly, we consider that particles are born around their guiding radius $R_g(t)$ at $t_i=-3.0$ Gyr. Indeed, $R_g(t)$ as defined in this work is comparable to an average position of a particle, filtering out its epicyclic oscillations. It is therefore a more reliable estimation of the birth position of the star. Newborn stars are also known for having nearly circular orbits \citep[see][]{Eggen1962, Yu2023}, hence the relevance to use $R_g(t)$ instead of a random $R$ position in the epicyclic oscillations of the particles for their birth radius. The simulations originally start at $t=-4.0$ Gyr, and the first pericentre passage of $G_{\rm SMC}$ happens at $t\sim-1.6$ Gyr. Therefore, even though some simulations (e.g., K4 and K6\textsubscript{i}) are not fully relaxed at $t_i$, this time stand as a good compromise to use as a stable morphology of $G_{\rm LMC}$. 
 
We therefore apply a typical disc galaxy's radial metallicity gradient of -0.04 dex/kpc to all stellar particles at $t_i$ and follow their mixing with respect to time. This radial metallicity profile is obtained in \cite{metallicitygradient} from a catalogue of disc galaxies in a hydrodynamical simulation and is checked consistent with observations. It allows to write the expression of the initial metallicity profile as follows:
\begin{equation}
    Z(t_i,R_g)=Z(t_i,0)\times10^{-0.04\times R_g}
    \label{eq:metalprofile}
\end{equation}
Where $Z(t_i,0)$ is the initial metallicity at the centre of $G_{\rm LMC}$ that we keep unknown in this work. Because of the unit of the metallicity gradient,  $R_g(t)$ must be given in kpc in this formula. We compare the evolving metallicity profile with the initial one at each snapshot $t$, normalised by $Z(t_i,0)$. For testing purposes, we checked different metallicity profile shapes (linear, squared...) and values for the gradient (up to 50\% stronger and weaker than observational data), and different values for $t_i$. None of these changes had a significant impact on the tendencies of our results, as long as the two following requirements are fulfilled. Firstly, the metallicity profile has to be decreasing, which is a very weak assumption confirmed by both theory and observations \citep[e.g.,][]{10.1093/mnras/stt1117}. Secondly, since pericentre passages of $G_{\rm SMC}$ strongly perturb the morphology and dynamics of $G_{\rm LMC}$, $t_i$ has to be  selected sufficiently far from such events to better represent a more relaxed, though not fully settled, galactic disc. As an output, this method allows to map $[Z(t,R_g)-Z(t_i,R_g)]/Z(t_i,0)$ with respect to $t$ and $R_g(t)$. These maps provide a quantification of the impact of tidal interactions on $G_{\rm LMC}$'s metallicity profile, and additionally, a direct visualisation of the radial mixing in the galactic disc. Indeed, opposite migrations yield zero net churning flux (see the previous paragraph) but still drive radial mixing, as outward migrators enrich the outer disc and inward ones dilute the inner regions. The metallicity maps are therefore complementary to the churning flux maps, enabling to quantify the flattening of the metallicity profile where net fluxes undergo a cancelling effect.

\begin{figure*}
    \centering
    \includegraphics[width=0.8\textwidth]{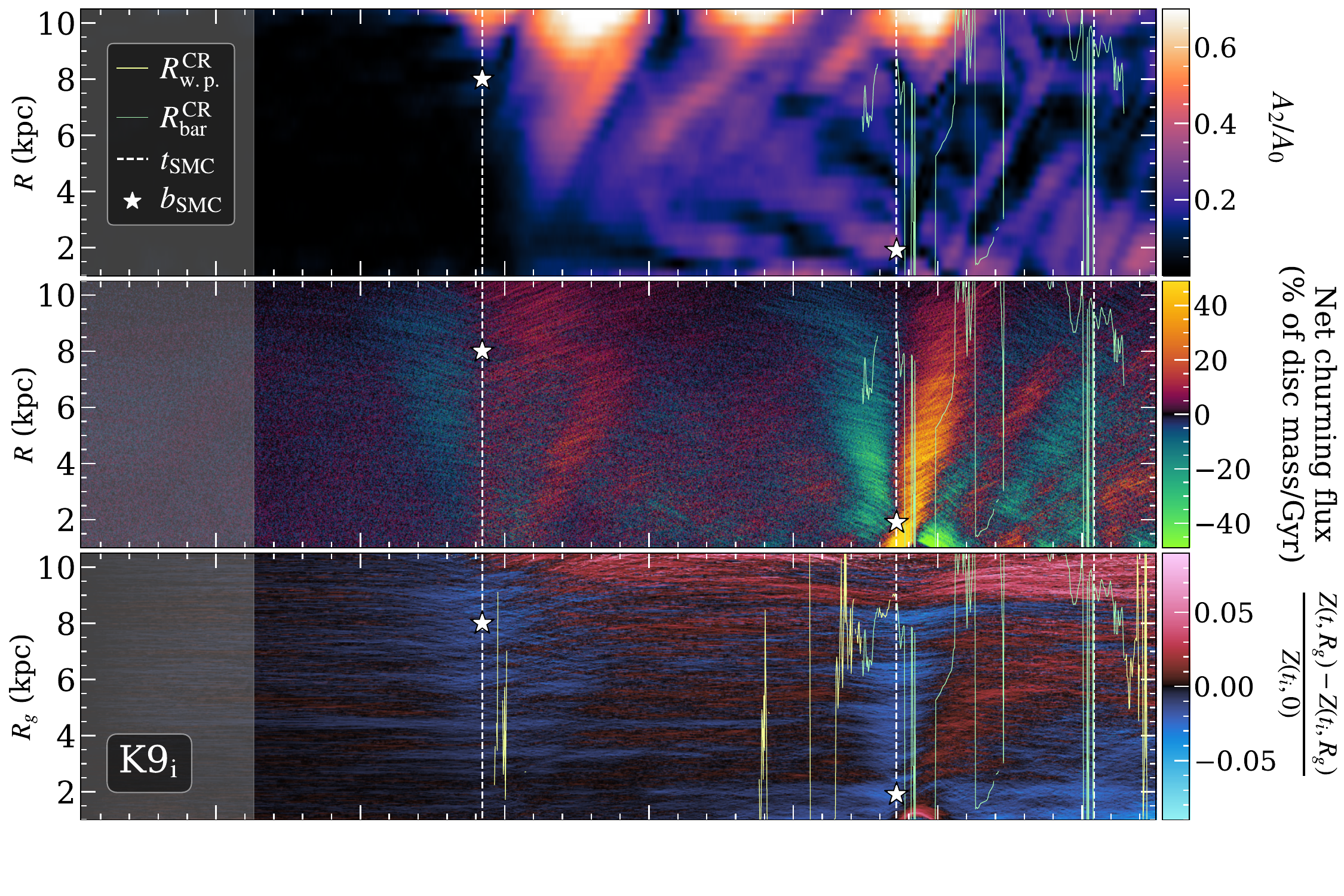}
    \includegraphics[width=0.8\textwidth]{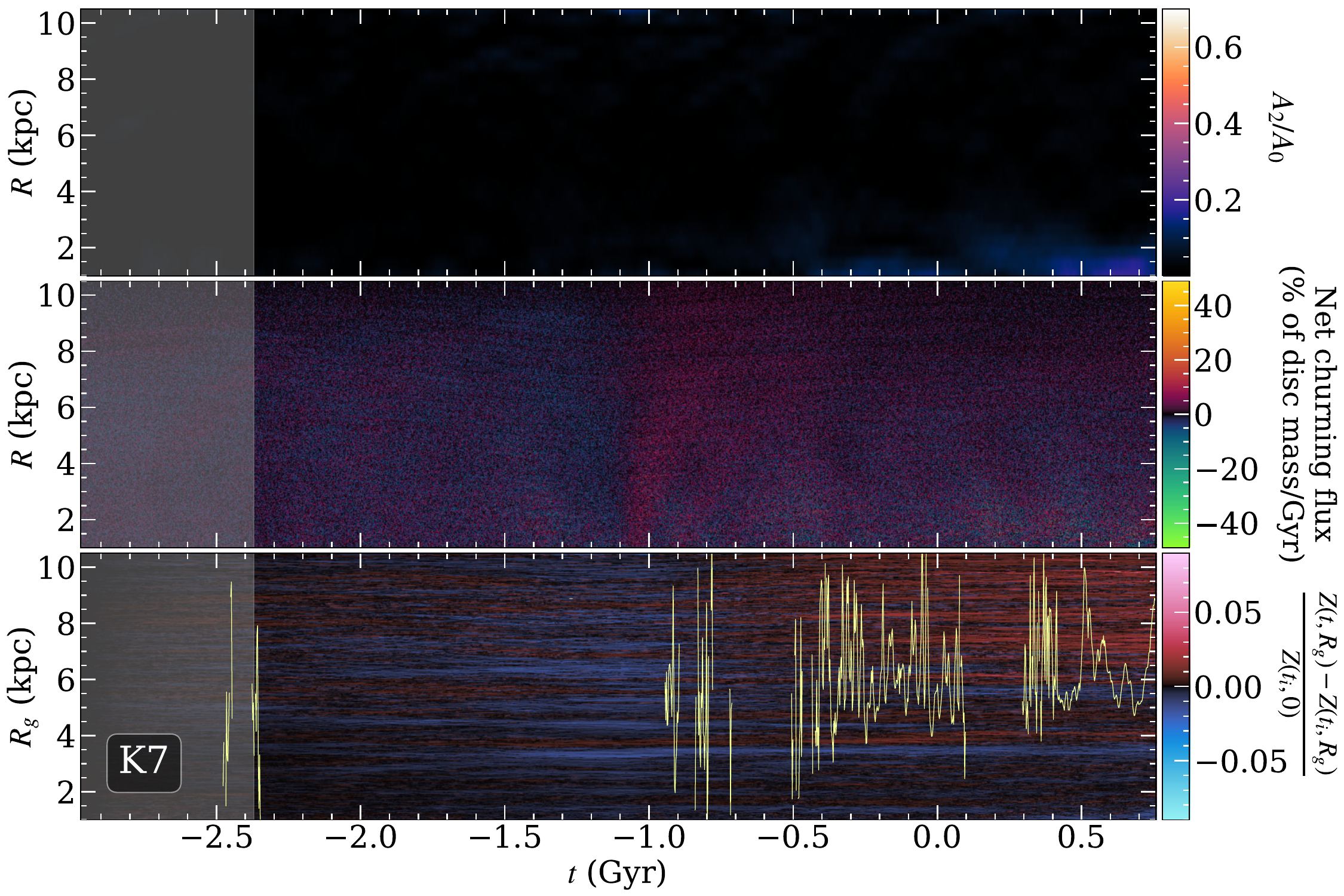}
    \caption{$A_2/A_0$ Fourier mode strength (top panel), difference between the churning fluxes outwards and inwards (central panel, counted positive outwards), and normalised metallicity evolution (bottom panel) maps with respect to time and galactocentric radius $R$ or guiding radius $R_g(t)$ for the K9\textsubscript{i} (interacting $Q=1.5$ simulation, top three panels) and K7 (isolated $Q=1.5$ simulation, bottom three panels) simulations. The grey area masks the first $0.6$ Gyr displayed, when the disc is not fully relaxed yet. For the K9\textsubscript{i} maps, the times of the pericentre passages of $G_{\rm SMC}$ are shown in white dashed lines, and its impact parameters are the white stars. The corotation radius is plotted in green and yellow for resonance with a bar or a weak central pattern, respectively.}
    \label{MAPS79}
\end{figure*}

Additionally, we display on each map the radius of corotation $R_{\rm CR}(t)$ with the central bisymmetric pattern. Corotation resonance is known to be able to trap stars in orbits close to $R_{\rm CR}(t)$, but also to act as a potential barrier, crossed only by stars with highly chaotic or high-energy orbits \citep[e.g.,][]{binney2011galactic,diffusionbar2025}. $R_{\rm CR}$ therefore is a widely studied parameter with a direct link with churning in isolated galaxy. However, other weaker resonances such as inner and outer Lindblad resonances are beyond the scope of this work. We estimate $R_{\rm CR}$ by solving the following implicit equation:
\begin{equation}
\langle \Omega \rangle (R=R_{\rm CR},t)-\Omega_p(t)=0
\label{eq:R_CR}
\end{equation}
Where $\langle \Omega \rangle(R,t)$ is obtained by averaging the rotation frequencies of all the particles within each of the radial bins used to map $A_2/A_0$ at each snapshot $t$. $\Omega_p$ is the rotation frequency of the central pattern presented in \citetalias{jimenez2024kratos} calculated with the Dehnen method (see \citeauthor{Dehnen_method}, \citeyear{Dehnen_method}, widely used in recent works as \citeauthor{semczuk2024pattern}, \citeyear{semczuk2024pattern}; \citeauthor{zhang2024deciphering}, \citeyear{zhang2024deciphering}; \citeauthor{2025jimenezbarstop}, \citeyear{2025jimenezbarstop}). A galaxy is considered barred if $A_2/A_0>0.2$ in its central region. If the latter condition is fulfilled, we refer to the corotation radius with the bar as $R^{\rm CR}_{\rm bar}$. On the other hand, when $G_{\rm LMC}$ does not feature a bar pattern ($A_2/A_0<0.2$ in the central region), it can still host a  non-axisymmetric rotating pattern with which stars might resonate. We hereafter refer to that sort of feature as a weak pattern (w.p.) and to its corotation radius as $R^{\rm CR}_{\rm w.p.}$ as displayed in Figs. \ref{MAPS79}, \ref{MAPS46}, and \ref{MAPS13}, only in metallicity maps for the sake of readability. For reference, we display the latter for all the snapshots with $0.1<A_2/A_0<0.2$. The condition $A_2/A_0>0.1$ being the minimum threshold allowing to compute $\Omega_p$ and therefore $R^{\rm CR}_{\rm w.p.}$ accurately.

\section{Results}
\label{sec:results}

Our main results describe the interplay between non-axisymmetric structures (Sect.~\ref{sub:Bisym}), radial migration (Sect.~\ref{sub:RM}), metallicity distribution (Sect.~\ref{sub:metal}) and corotation resonance (Sect.~\ref{sub:res}) under the influence of tidal interactions within $G_{\rm LMC}$ in the KRATOS simulations displayed on Table~\ref{table:KRATOS}.
 
We display our results in maps surveying $G_{\rm LMC}$'s interior with respect to time $t$ and galactocentric radius $R$ (or guiding radius $R_g(t)$), see Figs.~\ref{MAPS79} for K7 and K9\textsubscript{i}, \ref{MAPS46} for K4 and K6\textsubscript{i}, and~\ref{MAPS13} for K1 and K3\textsubscript{i}. These maps were obtained by applying the method described in Sect.~\ref{sub:methodMaps}. Each figure includes three maps of an interacting simulation (the three upper panels) and three maps of its isolated counterpart (the three lower panels). For each of the simulations, the first map is the $A_2/A_0$ map in the galactic disc, displaying the non-axisymmetric patterns strength. The second map displays the churning fluxes (or radial migration fluxes) in the galaxy and the last map shows their impact on the initial metallicity distribution. The latter metallicity map displays the radius of corotation with the bar and also with a weak non-axisymmetric pattern, whereas only corotation with the bar is displayed on the other maps. Additionally, all interacting maps feature white vertical dashed lines and white stars matching the pericentre passages of $G_{\rm SMC}$ and its impact parameters on $G_{\rm LMC}$'s disc, respectively.

\subsection{Non-axisymmetric patterns}
\label{sub:Bisym}
The correlation between the presence of non-axisymmetric structures and the tidal interactions was already established in the KRATOS presentation paper \citepalias{jimenez2024kratos}. However, their results were obtained by calculating the median of $A_2/A_0$ on the entire bar region. Contrastingly, our work allows to extend the study to the outer disc, and visualise $A_2/A_0$ on a much thinner binning, providing more details. We present our results for both isolated $G_{\rm LMC}$ and interacting systems.
 
We first present the isolated simulations: K7, K4 and K1 (see the three bottom panels of  Figs.~\ref{MAPS79}, \ref{MAPS46} and~\ref{MAPS13}, respectively). Being free from tidal forces with foreign systems, $G_{\rm LMC}$ only depends on its secular evolution to form non-axisymmetric patterns. None of these simulations show strong spiral arms at the relaxed state \citepalias[see Fig.~3 of][for a visualisation of the galaxies at $t=0$]{jimenez2024kratos}. Nonetheless, because their difference in the Toomre parameter $Q$, they present dissimilar morphological evolutions. The most stable of them (K7, with $Q=1.5$) is unable to form a strong bar within the total simulated 4.68 Gyr, only a very weak central pattern which is slowly growing from $t\sim-1.0$ Gyr until the end of the simulation. Additionally, K7 does not form any spiral arm during its evolution, resulting in a very dark  $A_2/A_0$ map (i.e. with a very low amplitude) in Fig.~\ref{MAPS79}. Contrastingly, the most unstable isolated simulation (K4, with $Q=1.0$) is able to form, maintain and grow a strong bar pattern longer than 2 kpc from $t\gtrsim-2$ Gyr \citepalias[also described in][]{jimenez2024kratos}. Additionally, it sparsely showcases small and local spiral arms (i.e. not covering the entire disc, some stains can be seen in K4's $A_2/A_0$ map of Fig.~\ref{MAPS46}). In between both simulations, K1 ($Q=1.2$) develops a weak pattern in its centre from $t\gtrsim-1.5$ Gyr growing into a strong but very variable bar at $t\gtrsim1.0$ Gyr. This variability can be seen in the low $R$ range of Fig.~\ref{MAPS13}'s $A_2/A_0$ map, in particular at the end of the simulation where the upper limit of the strong $A_2/A_0$ region oscillates. Furthermore, this simulation presents spiral patterns similar to K4's, but in a lesser extent and weaker strength.
 
Secondly, we present the interacting simulations: K9\textsubscript{i}, K6\textsubscript{i} and K3\textsubscript{i}, see the three top panels of  Figs.~\ref{MAPS79}, \ref{MAPS46} and \ref{MAPS13}, respectively. Despite sharing the same Toomre parameter $Q$ as their isolated counterparts (K7, K4, and K1, respectively), the interacting galaxies exhibit marked morphological differences relative to them. These unlikenesses triggered by tidal interactions may include both formation and destruction of bar patterns, and creation of strong global spirals with a large variety of shapes. The two latter results being already explicitly presented in \citetalias{jimenez2024kratos}, the novelty of this work is to allow to see the evolution of their $A_2/A_0$'s radial distribution. The spiral pattern evolution of $G_{\rm LMC}$ is similar in all interacting simulations. Namely, $G_{\rm LMC}$'s outer disc takes a global spiral shape after the first pericentre passage of $G_{\rm SMC}$. The high $A_2/A_0$ area corresponding to this spiral shape appears to propagate outwards in all the interacting simulation's $A_2/A_0$ maps. Equivalently, the low $A_2/A_0$ areas follow the same direction. We mention that the first interaction is also the longest-range pericentre passage of $G_{\rm LMC}$, and the only one with an impact parameter in the outer disc. Furthermore, these spiral arms remain short-lived and variable, being susceptible to be destroyed and created again. They also are much weakened after the second galactic interaction.
 
Although these previously listed phenomena are similar for all the interacting simulations, $G_{\rm LMC}$'s bar pattern evolves differently depending on the Toomre parameter $Q$. For instance, in K9\textsubscript{i} ($Q=1.5$), $G_{\rm LMC}$'s bar presents a very singular behaviour. It is formed during the first pericentre passage of  $G_{\rm SMC}$ (the first white vertical dashed line of Fig.~\ref{MAPS79}), initially shorter than $1$ kpc and circled by a ring. This bar is not well detected by the Dehnen method but keeps its shape until the second galactic encounter. During this encounter, the bar almost stops rotating for around 100 Myr \citep[see][]{2025jimenezbarstop}, shifts from the centre and gets rid of its ring. After this encounter, it will relax and continue growing until the end of the simulation. On the other hand,  K6\textsubscript{i} ($Q=1$, see Fig.~\ref{MAPS46}), as its isolated counterpart K4, starts forming a strong steady bar at $t\sim-2$ Gyr (i.e. before the first interaction). This bar grows during all the simulation, ultimately being stronger and larger than K4's bar. However, because of the tidal interactions, K6\textsubscript{i}'s bar is more variable, impacting its corotation radius (see Sect.~\ref{sub:res}). Finally, K3\textsubscript{i}'s bar ($Q=1.2$, see Fig.~\ref{MAPS13}) presents a similar behaviour as K6\textsubscript{i}'s. The main difference is that the bar creation starts during the first pericentre passage (i.e. after K6\textsubscript{i}'s bar). It therefore starts slightly before K1's bar (being K3\textsubscript{i}'s isolated counterpart).
 
The morphology of $G_{\rm LMC}$ at a given time and radius is inherently connected to the motion of its stars. Consequently, $A_2/A_0$ maps provide complementary informations to churning fluxes and metallicity distribution maps presented in the next sections.

\subsection{Radial Migration}
\label{sub:RM}
The general tendencies of radial migration are described by the churning fluxes in $G_{\rm LMC}$'s disc (see Sect.~\ref{sub:methodMaps}). These fluxes are represented in the second panels of each group of maps in Figs.~\ref{MAPS79}, \ref{MAPS46} and \ref{MAPS13}. Although some simulations can spontaneously show some exceptions (see the end of this subsection), the general tendencies are similar within each group of simulations: isolated and interacting.
 
The isolated simulations do not display strong global churning fluxes, usually not exceeding $\sim$15-20\% of disc mass per Gyr in any direction. There are still non-zero fluxes in the galactic discs, especially in the least stable simulations (K4 and K1, see Figs.~\ref{MAPS46} and \ref{MAPS13}, respectively). These fluxes appear to be oscillating with time in the centre of the galaxy—especially at the end of the simulations. On the other hand, the churning fluxes are similar to a diffuse noise in the outer disc, with some spontaneous larger patterns – in particular when the galaxy is not fully relaxed. These features are also present in the interacting simulations, hence the relevance of this isolated control group.
 
The interacting simulations' churning fluxes showcase a strong global wave pattern reaching a first minimum at a time perfectly matching the pericentre passages of $G_{\rm SMC}$. The churning fluxes decrease before the pericentre passages, then increase back with a similar amplitude, peaking around the impact parameter. Their values reach $\sim50\%$ of disc mass per Gyr in both directions during the strongest interaction: the second pericentre passage. This radial migration wave continues to oscillate with a shorter wavelength after the pericentre passages. Although it is most notably visible in K9\textsubscript{i}'s second interaction because of the high stability of $G_{\rm LMC}$ in this simulation (see second panel of Fig.~\ref{MAPS79}), this wave pattern exists and is visible in all the interacting simulations. It covers the entire galactic disc and oscillates faster near the galactic centre than in the outer disc (i.e. the wavelength is shorter close to the centre, hence the appearance of churning fluxes in diagonal bands on the maps). The latter statement remains valid when the impact parameter of $G_{\rm SMC}$ is in the outer disc (see for instance the first pericentre passage in K9\textsubscript{i} in Fig.~\ref{MAPS79}).
 
Some exceptions to the latter statements still lie in the simulations. One of the most prominent can be seen in K7 at $t\sim-1.1$ Gyr (see the second last panel of Fig.~\ref{MAPS79}) where a weak wave pattern similar to the ones happening during tidal interactions is triggered. We discuss this phenomenon in Sects.~\ref{sub:res} and \ref{sub:isolated}. This behaviour may also appear in the other simulations, though it is less discernible due its very low amplitude versus the fluxes created by their instabilities.

\subsection{Metallicities}
\label{sub:metal}

As mentioned in Sect.~\ref{sub:methodMaps}, the study of the radial metallicity distribution allows both to link radial migration in $G_{\rm LMC}$'s disc with an observable, and provide complementary informations on the radial mixing induced by churning. We again present the results for each group of simulations.
 
In the isolated group, churning tends to enhance the metallicity in the outer disc, and reduce it closer to the centre. It therefore directly weakens the metallicity gradient with time. This effect is much amplified for low $Q$ simulations because of their instability (see K4 and K1 in third panels of Figs.~\ref{MAPS46} and \ref{MAPS13}, respectively). This metallicity difference can reach $\sim$7-8\% of the initial maximum metallicity $Z(t_i,0)$ in K4 and K1, and $\sim$3\% of $Z(t_i,0)$ in K7. Moreover, a lower Toomre parameter $Q$ correlates with a sharper demarcation between regions of enhanced and reduced metallicity. We hereafter refer to these regions of unclear demarcation as ``mixing zones'' (see for instance the $3<R_g<6$ kpc region in almost all of the K1 simulation in the last panel of Fig.~\ref{MAPS13}). Finally, lower values of $Q$ are associated with a more rapid development of these effects.
 
The aforementioned effects are also applicable to the interacting group. However, they present some differences due to the tidal interactions. The most visible of which is a metallicity drop of $\sim$3-5\% of $Z(t_i,0)$ during the two first pericentre passages of $G_{\rm SMC}$—therefore matching with the minimum of the wave presented in Sect.~\ref{sub:RM}. This drop is most distinctly visible in K9\textsubscript{i} (see third panel of Fig.~\ref{MAPS79}), but also appears in K6\textsubscript{i} and K3\textsubscript{i} (see third panels of Figs.~\ref{MAPS46} and \ref{MAPS13}, respectively). After this drop, K9\textsubscript{i} features a sudden metallicity increase that propagates outwards across the disc, before decreasing again. This effect is discussed further in Sect.~\ref{sec:discussion}. Finally, tidal interactions tend to strengthen the radial mixing presented for isolated simulations. For instance, the radial metallicity gradient is weaker at the end of K9\textsubscript{i} than it is for its isolated counterpart K7 at the same time. 
 
We emphasise that these changes in metallicity do not describe blurring but solely churning. Indeed, they depend only on the guiding radius $R_g(t)$ and are therefore a metric to quantify the radial migration in the disc. To conclude, the effects presented in this subsection can be seen as the typical impacts of radial migration—with or without tidal interactions—on the metallicity distribution within the galaxy.

\subsection{Influence of the corotation resonance}
\label{sub:res}
As noted in Sect.~\ref{sub:methodMaps}, the corotation radius $R_{\rm CR}$, namely, the radius at which the corotation resonance lies, is directly linked with the pattern speed and strength of the central non-axisymmetric pattern (a bar when $A_2/A_0>0.2$ in that region, otherwise a weaker pattern). Thus, the results presented in this section are to be linked with those mentioned in Sect.~\ref{sub:Bisym}, where we shown that the non-axisymmetric patterns are highly variable in most of the KRATOS simulations. As a consequence, the corotation radius is also very variable, and is largely affected by tidal interactions. It is, however, still a good indicator of a radius acting both like a trap and a barrier, especially for the isolated simulations where $G_{\rm LMC}$ hosts a bar.
 
Due to the high variability of $R_{\rm CR}$ in interacting simulations, we focus on the results obtained in the isolated simulations. Our main results can be observed in K4 and K1, because of the strength and stability of their bars. First, in K4, $R_{\rm CR}$ is fairly stable and almost always lies at the demarcation between enhanced and reduced metallicities (i.e. in the small mixing zone, see Fig.~\ref{MAPS46}). Second, in K1, $R_{\rm CR}$ is more variable, and is slightly higher than the radius of the mixing zone (see Fig.~\ref{MAPS13}). In both of these simulations, the amplitude of the variations of $R_{\rm CR}$ matches the radial width of the mixing zone. This correlation between metallicities and $R_{\rm CR}$ is discussed in Sect.~\ref{sec:discussion}. Furthermore, we stated in Sect.~\ref{sub:RM} that the churning fluxes behaved differently closer to the centre than in the outer disc; rapidly oscillating with time near the centre and showing larger noisy patterns in the outer disc. We find that in both K4 and K1, $R_{\rm CR}$ appears to be the demarcation between these two behaviours. The two latter statements could also be applied to K6\textsubscript{i}, even though the computation of $R_{\rm CR}$ is complexioned by the interactions.
 
Even though K7 does never form a strong bar, $R_{\rm CR}$ suddenly increases at $t\sim-0.9$ Gyr, suggesting that a weak non-axisymmetric pattern is growing. The time of this growth almost matches the triggering of the weak radial migration wave mentioned in Sect.~\ref{sub:RM}. For reference, we mention that in K9\textsubscript{i}, $R_{\rm CR}\sim11$ kpc between $t\sim0$ and $t\sim0.5$ Gyr, due to $G_{\rm LMC}$'s slowly rotating bar. This high $R_{\rm CR}$ correlates with the sudden decrease of metallicity in the same time span at $R_g\sim8.5$ kpc (see Fig.~\ref{MAPS79}). The link of causality between a high $R_{\rm CR}$ and a sudden drop of metallicity below it remains uncertain. We find a similar behaviour, though in a lower extent, in K3\textsubscript{i} and K7 (at $t\sim-0.7$ Gyr in the third panel of Fig.~\ref{MAPS13}, and at $t\sim0.5$ Gyr in the last panel of Fig.~\ref{MAPS79}, respectively).

\section{Discussion}
\label{sec:discussion}

\subsection{Isolated simulations}
\label{sub:isolated}
Isolated simulations (K1, K4, K7) provide a control group to study tidal interactions. Yet, they display intrinsic dynamics that merit analysis, and allow comparison with prior works.

A lower Toomre parameter $Q$ \citep{1964ApJ...139.1217T} increases instabilities and non-axisymmetric structure in $G_{\rm LMC}$, and in all isolated simulations radial migration correlates with these patterns \citep[which is consistent with e.g.,][]{toomrebar,toomrespiral}. This highlights a clear connection observed in all isolated simulations: radial migration strongly correlates with the presence of non-axisymmetric patterns. Thus, we observe that a lower $Q$—therefore a stronger non-axisymmetric component—leads to increased churning fluxes (see Sect.~\ref{sub:RM}) accompanied by more rapid and intense radial mixing (i.e. weakening of the metallicity gradient, see Sect.~\ref{sub:metal}). This tendency is at the basis of the concept of radial migration and has always been identified as its main cause in isolated galaxies. Our results therefore are strongly consistent with those found in the existing literature of the field \citep[e.g.,][]{sellwood2002radial,schonrich2009chemical,iles2024impact,marques2025bar}. Furthermore, all simulations exhibit noise-like, incoherent churning throughout (including initial relaxation; grey area in Figs.~\ref{MAPS79}, \ref{MAPS46}, \ref{MAPS13}). Given our metallicity definition (dependent on $R_g(t_i)$; see Sect.~\ref{sub:methodMaps}), this supports radial migration as the primary driver of metallicity smoothing, especially in the least stable galaxies.

As described in Sect.~\ref{sub:RM}, the churning fluxes behave differently in the outer disc than close to the centre of $G_{\rm LMC}$. Indeed, the fluxes show almost coherent oscillations near the galactic centre whereas they display noise-like patterns varying less rapidly in the outer disc. This feature is especially visible in barred systems, partly because the fluxes themselves are stronger due to their less stable nature (see for example the second last maps of Figs.~\ref{MAPS46} and \ref{MAPS13}). For instance, the period of these oscillations after $t\sim0.5$ Gyr near the centre of $G_{\rm LMC}$'s centre in K1 ($\sim$ four full oscillations per Gyr) matches double of the bar rotation period in this simulation. Furthermore, these oscillations are clearer when the bar is stronger (i.e. when $R_{\rm CR}$ is displayed with a green line in Fig.~\ref{MAPS13}). To summarise, these oscillations occur each time a strong bar undergoes half a rotation in K1, they therefore may be a consequence of the local changes in the gravitational potential created by the bar's motion. Thus, it strengthens the link between non-axisymmetric patterns and radial migration. This reasoning can be applied to the other simulations, but is harder to quantify. Indeed, K4's churning fluxes map features numerous local patterns that may be triggered by instabilities due to its critical value of $Q$, slightly hiding and perturbing these central oscillations. Regarding K7, it only features very low fluxes, and quantifying the period of these oscillations would be challenging. We still mention that small oscillating patterns are visible in K4, and, in a lesser extent, in K7 near the end of the simulation when the weak non-axisymmetric pattern is formed in its centre. Additionally, it appears in both K4 and K1 that the boundary between the central oscillation-dominated zone and the outer disc is $R_{\rm CR}$. This is consistent with the fact that $R_{\rm CR}$ is known to act as a potential barrier \citep[e.g.,][]{binney2011galactic,diffusionbar2025}, it therefore also serves as a barrier between the different regimes of radial migration.

There is a strong correlation between the radial width of the mixing zone (as defined in Sect.~\ref{sub:methodMaps}) and the amplitude of $R_{\rm CR}$'s rapid pseudo-oscillations. This effect is particularly visible in K4 and K1 which feature a stable $R_{\rm CR}$ with a very small mixing zone and an oscillating $R_{\rm CR}$ with a wider mixing zone, respectively. This correlation could be explained by the fact that the corotation radius tends to trap particles in its radial range. Thus, when $R_{\rm CR}$ oscillates with some amplitude, stars trapped in corotation mix together creating this mixing zone which width matches this amplitude. Although the most reliable occurrences of this correlation happen when $R_{\rm CR}$ is calculated with a strong bar pattern, we mention that a similar effect is visible for a weaker pattern in K1 before $t\sim-1$ Gyr. Indeed, in this time span, K1's $R_{\rm CR}$ is strongly variable, correlating with a mixing zone covering almost the entire galactic disc with metallicity enhancements near the centre and reductions in the outer disc (see last panel of Fig.~\ref{MAPS13}). The mixing zone width tends to reduce when the bar grows stronger and when $R_{\rm CR}$ is less strongly varying. A similar tendency is visible in K7 after $t\sim-0.5$ Gyr, but because the mixing zone is less distinguishable and the central pattern is very weak in this simulation, we refrain from drawing strong conclusions on this effect in K7.
 
As mentioned in Sect.~\ref{sub:RM}, K7's radial migration hosts a major peculiarity at $t\sim-1.1$ Gyr (see K7's churning flux map in Fig.~\ref{MAPS79}). Indeed, a small radial migration wave seems to be triggered at that time (that kind of wave is discussed more extensively in Sect.~\ref{sub:interacting}), correlating with a weak  metallicity drop (< 1\% of $Z(t_i,0)$) in the inner disc. A few tenths of Myr after the wave is triggered, a weak non-axisymmetric central pattern is formed in $G_{\rm LMC}$'s centre. These two events are likely linked because bars are created by instabilities known for being a cause of radial migration \citep[e.g.,][]{barform2014, barform2024}. Moreover, bar formation is often considered as the strongest cause of churning in isolated galaxies \citep[e.g.,][]{halle2015quantifying}. The churning flux wave triggered shortly before the appearance of a weak bar-like pattern in K7 is therefore in agreement with the scientific literature preceding this work, and we conclude that it may be triggered by the bar forming instability. It is also possible that K7 could develop a stronger bar over a longer timescale than that covered in the simulation, as we do not observe its weak non-axisymmetric pattern ceasing to grow. Similar churning patterns are slightly visible in K4 and K1 at similar times, they are less visible because of their higher background churning flux intensity. It is therefore not obvious to know if the bar formation is at their origin too, but we do not exclude this possibility. We also note the bar buckling instability, which arises after bar formation, reshapes the vertical structure into an ``X'' or peanut shape \citep{bucklinginst, buckling2}. This instability can alter bar strength and pattern speed, shifting resonances and releasing stars previously trapped in resonant orbits, thereby punctually impacting radial migration \citep[e.g.,][]{halle2018thickdisc, khoperskov2020escapees}. It is therefore another instability expected to generate coherent fluxes in K4 and K1.

Overall, our results with isolated simulations reproduce established radial-migration behaviours, providing a baseline for the interacting cases and reinforcing the robustness of our subsequent conclusions.

\subsection{Interacting simulations}
\label{sub:interacting}

The interacting group of simulations (K3\textsubscript{i}, K6\textsubscript{i} and K9\textsubscript{i}) hosts the main results of this work and therefore deserve an extensive discussion. In this section, we focus on the results appearing as a consequence to the tidal interactions. Indeed, we expect that most of the results discussed for the isolated simulations (see Sect.~\ref{sub:isolated}) are also applicable to their interacting equivalents, although usually not directly visible because hidden behind the stronger effects we discuss in this section. Thus, the goal of this section is to provide an interpretation of the effects of tidal interactions on galaxy morphology, radial migration, radial metallicity distribution, and the interplay between all of them.
 
In all interacting simulations, the first tidal interaction creates a spiral structure remaining ubiquitous in $G_{\rm LMC}$ until the next pericentre passage of $G_{\rm SMC}$. Two main reasons are to be linked with the creation of this structure. First, the impact parameter ($b_{\rm SMC}\sim$ 8.0 kpc) means that $G_{\rm SMC}$ is located near the outer disc of $G_{\rm LMC}$ during this interaction. Therefore, the spiral structure is created where the tidal interaction is the strongest. Second, this is the longest-ranged pericentre passage of $G_{\rm SMC}$ with a total distance of 11.8 kpc between the galaxies and a vertical distance of $z_ \text{peri} = -8.7$ kpc whereas $z_\text{peri}\sim0$ for the two later passages. To summarise, the only interaction that can be seen creating a strong large scale spiral structure is the longest-range interaction and has an impact parameter in the outer disc. We also note that a similar churning flux wave pattern is triggered during both first interactions, but only the first is seen creating global spiral arms. However, during the first interaction, the churning wave is triggered before the formation of spiral arms suggesting that the non-axisymmetric pattern is a consequence of radial migration (see the radial migration fluxes in green and orange around the first white vertical dashed line, both appearing before the orange overdensity of $A_2/A_0$ in Figs.~\ref{MAPS79}, \ref{MAPS46}, and \ref{MAPS13}). In other words, the creation of large scale spiral arms in the KRATOS simulations is not merely a consequence of a change in the stars' position, but a radial shift in their guiding centre. The latter result is consistent with the pre-established link between radial migration and transient spiral arms \citep[e.g.,][]{sellwood2002radial, solway2012radial, marques2025bar}.
 
A radial migration wave triggered by the two first interactions between $G_{\rm LMC}$ and $G_{\rm SMC}$ is visible in all interacting simulations' churning flux maps (see Figs.~\ref{MAPS79}, \ref{MAPS46}, \ref{MAPS13}). These waves systematically show inward fluxes, then a minimum matching with the pericentre passage and outward fluxes, eventually showing more oscillations in the second and stronger tidal interaction. As mentioned in Sect.~\ref{sub:metal}, all pericentre passages match with a metallicity drop of a few percent of $Z(t_i,0)$ in almost all the galactic disc. This drop is a direct consequence of the radial migration wave. Indeed, the initial metallicity profile is decreasing with $R_g(t)$. Therefore, when the net churning fluxes describe a general inwards migration, regions that were initially more metal-rich become populated by particles that were initially less metal-rich, leading to a drop in metallicity in those regions. Furthermore, during the second interaction in K9\textsubscript{i}, metallicity increases again following this drop and mirrors the shape of the churning flux wave—peaking after the outward fluxes (see the third panel of Fig.~\ref{MAPS79}). This reversal further reinforces the causal link between radial migration and variations in metallicity. We also mention that these metallicity drops and increases are just redistributions of the metallicity in the disc, because star formation and evolution were overlooked (see Sect.~\ref{sub:methodMaps}), the total amount of metals in the model remains unchanged.
 
In all of the interacting simulations, the churning wave starts with inward fluxes and is followed by a bouncing effect creating slightly stronger outward migration. We expect the order of events to be due to the nature of tidal interactions. Indeed, in the KRATOS simulations, $G_{\rm SMC}$ follows an almost polar orbit around $G_{\rm LMC}$. This implies that the tidal forces apply in-plane components towards the centre of $G_{\rm LMC}$, and normal components tending to thicken it. Thus, the in-plane component is in agreement with the first direction of the churning fluxes and the normal component explains the increase in $G_{\rm LMC}$'s disc scale height after the interactions \citepalias[see Fig. 6 of][]{jimenez2024kratos}. We attribute the subsequent outward fluxes to the relaxation of particles that gained energy during the interaction. In fact, the disc height reduces right after the pericentre passages—especially the first—thus meaning that stars lose some of the vertical kinetic energy they gained after the interaction. Because the stars are dragged towards $G_{\rm LMC}$'s centre, therefore with a radial component, some of the vertical kinetic energy of the stars can be converted into azimuthal kinetic energy which increases their $R_g(t)$ driving the outward churning fluxes. The in-depth link between radial migration and vertical motion remains unclear and is out of the scope of this work. Indeed, while some works show that the strongest migrators are the particles with the least vertical motion \citep[e.g.,][]{vera2014effect,vera2016imprint}, more recent studies present other regimes where the vertical motion has less or no impact on the radial migration \citep[e.g.,][]{halle2018thickdisc,mikkola2020radial}. Finally, we mention that the churning fluxes oscillate increasingly faster over time toward the galactic centre. This may be attributed to the shorter dynamical timescales governing orbital evolution near the galactic centre.
 
The estimation of $R_{\rm CR}$ in interacting simulations is challenging; we therefore treat our results with caution. The corotation radius exhibits abrupt changes and even discontinuities in the interacting simulations, especially near the pericentre passages of $G_{\rm SMC}$. We attribute these irregularities to the unpredictable nature of morphological changes in the galaxy, the sudden changes in the bar pattern speed, the off-centring of the bar, and the asymmetries induced by tidal interactions \citep[see][]{MarieAsymmetries25}. All of these parameters may interfere with the measurement of the bar pattern speed and the estimation of $G_{\rm LMC}$'s rotation curve, both of which appearing in $R_{\rm CR}$'s calculation (see Sect.~\ref{sub:methodMaps}). However, we still notice some local recognisable correlations between the resonance with a strong bar and local patterns in the metallicity maps. Indeed, as described in Sect.~\ref{sub:res}, K3\textsubscript{i} and K9\textsubscript{i} exhibit local metallicity drops in the outer disc (see the blue horizontal plumes in the third panels of Fig.~\ref{MAPS79} at $t\sim0.3$ Gyr and Fig.~\ref{MAPS13} at $t\sim-1.0$ Gyr). Both of these oddities happen at $\sim t_{\rm SMC}+500$ Myrs (i.e. after a $G_{\rm LMC}$-$G_{\rm SMC}$ pericentre passage) and correlate with the highest values of $R_{\rm CR}$ with a strong bar in their corresponding host simulations. The absence of both metallicity drop plumes in the outer disc and very high $R_{\rm CR}$ values in K6\textsubscript{i} further supports our interpretation of this correlation. Although the physical mechanism underlying this interplay remains uncertain, we remind that $R_{\rm CR}$ is able to drag and trap particles in its orbital range. Because both of the described local metallicity drops start during a tidal interaction, itself creating a larger scale metallicity drop, it is plausible that $R_{\rm CR}$ may contribute to maintaining the particles' positions in the outer disc as they were during the radial migration wave. Moreover, we do not draw strong conclusions on the effect of resonance with weak patterns in interacting simulations because of their high variability, and therefore only show their corotation radius in the metallicity maps for reference. Ultimately, we considered only the corotation with the bar and not the spiral arms as these lie outside the scope of this work, alongside the bar's inner and outer Lindblad resonances.
 
Regarding the long-term outcome of the tidal interactions, we note from the metallicity maps that radial mixing is stronger at the end of the interacting simulations compared to their isolated counterparts. Although this effect is most evident in the K7 and K9\textsubscript{i} pair (see Fig.~\ref{MAPS79}), that tendency exists in all pairs of simulations with at least a 1\% stronger metallicity increase in the interacting $G_{\rm LMC}$'s outer discs. This result is to be expected: tidal interactions have long been known for inducing dynamical disc heating \citep[e.g.,][]{tidalheatingold, gnedin2003tidal}. Thus, a disc that has experienced tidal interactions contains increased random motion, leading to enhanced radial mixing. Eventually, it implies a weaker metallicity gradient.
 
The results we presented for the impact of radial migration on the metallicity distribution may extend to other observable quantities linked to the stars and presenting a gradient. Indeed, given our approach to estimate the weakening of the metallicity gradient (see Sect.~\ref{sub:methodMaps}), our findings can be applied to any quantity that shows a decreasing (or, conversely, increasing—yielding mirrored results) trend with guiding radius $R_g(t)$. It also requires the assumption that it remains constant over time for a given star to hold reasonably well. Thus, because of radial migration, we expect a smoothing of the radial stellar age profile and the colour gradients in the galaxy with similar trends than what we observe for metallicity. Nevertheless, a more realistic quantification of these effects remains only a future prospect of this work and could require more sophisticated simulations.
 
In the end, the interacting simulations show very similar results during $G_{\rm SMC}$'s pericentre passages with a similar wave triggered in K9\textsubscript{i}, K6\textsubscript{i} and K3\textsubscript{i} during all interactions. Thus, we conclude that close-range tidal interactions can reshape the galaxy and dominate the dynamics of LMC-like galaxies from $Q=1.0$ to $Q=1.5$. Currently, models studying galactic dynamics and radial migration usually consider isolated galaxies \citep[e.g][]{halle2015quantifying, halle2018thickdisc, nagy2025comparing, marques2025bar} while galactic encounters are known to be common in our universe \citep[e.g.,][]{barnes2001merger,10.1093/mnras/278.3.727, antoja2018dynamically, merrow2024did}. We present results showing that they can be the dominant radial migration-driving mechanism in galactic discs, being able to generate churning fluxes of $\sim40\%$ of disc mass per Gyr. Thus, our work suggests that tidal interactions cannot be neglected to study galactic evolution in a realistic environment \citep[see][for a similar conclusion using other methods and scales]{Carr2022}.

\subsection{Limitations and future prospects}

We remain still far from fully understanding the principles underlying the impact of tidal interactions on galactic discs. Indeed, this work overlooks some key mechanisms that would require to be implemented for a more realistic study of radial migration. Those are listed below:

\begin{enumerate}
    \item The most obvious caveat of the KRATOS suite is the absence of hydrodynamics, and hence stellar formation and evolution, sacrificed for a higher resolution. Because stars are usually formed in nearly circular orbits \citep[e.g.,][]{Eggen1962, Yu2023}, they would populate almost inexistent orbital types in KRATOS and may have an impact on the overall migration. Furthermore, due to the slower rotation rate of the gas component, it would apply some damping to the stars' motion that was neglected in this work. Most importantly, incorporating star formation and stellar evolution would provide a far more accurate estimation of key observables of the galaxy. It would free this work from many of the assumptions we relied on to derive an estimate of the metallicity profile's evolution. Thus, the authors of this work are already developing new simulations that will study the effect of adding hydrodynamics, star formation and evolution physics, as a continuation of this project aiming to achieve more sophisticated and comprehensive simulations.
    \item As discussed in Sect.~\ref{sub:methodMaps}, all of our maps and most of our results are represented with respect to $R$ or $R_g(t)$ and $t$. However, it is known that both LMC and its simulated counterparts exhibit asymmetries \citep[e.g.,][]{luri2021gaia,jimenez2023kinematic, jimenezarranz2025verticalstructurekinematicslmc}, that we overlooked in this work. Thus, a finer description of migration accounting for variations both with the radius and the azimuthal angle across the disc, would be highly beneficial. This further study might be able to show that the impact position in the galactic plane correlates—or not—with the highest local churning fluxes. Some work has been done to quantify the asymmetries in the LMC disc \citep{MarieAsymmetries25} and a clear future prospect of this study would be to combine their results and methods with ours, enabling an investigation of the correlations between radial migration and asymmetries.
    \item We only presented results of the estimation of the bar or central pattern's corotation radius, therefore lacking an estimation of the Lindblad resonances and resonances with the spiral arms. Determining the precise locations of the resonances under the influence of tidal interactions is not straightforward, yet it is an aspect that more advanced hypothetical future studies should take into consideration. However, such an endeavour may require simulating a larger galaxy than $G_{\rm LMC}$ as the corotation resonance already reach the outer disc in some interacting simulations, therefore increasing the computational cost for the same resolution. This issue would be more pronounced when considering the outer Lindblad resonance or corotation with the spiral arms, with the risk of these radii exceeding the size of the galaxy itself.
    \item Because the only varying input parameters of this work are the Toomre parameter $Q$ and the presence of interactions, this study could be extended to additional simulations exploring the impact of other parameters. For instance, how is radial migration influenced by changes in the mass or shape of the dark matter halo? How would our results be affected by changes in the orbit of $G_{\rm SMC}$? Would the presence of a supermassive black hole at the galactic centre alter radial migration or reduce the impact of tidal interactions? Do tidal interactions produce similar effects in elliptical or irregular galaxies? Such further research could offer insights into regimes less dominated by tidal interactions than those explored in the present study. In particular, the impact of $G_{\rm SMC}$'s orbit and $G_{\rm LMC}$'s halo could be investigated using other KRATOS simulations as a natural continuation of this work.
\end{enumerate}

By leveraging simulations and observations, future studies could better characterize tidally-induced radial migration waves, which drive inward migration before pericentre passages and outward migration afterward. They may lead to enhanced surface brightness in the inner disc and reduced brightness in the outer disc during pericentre passages, along with the metallicity drop of $\sim$3-5\% of the maximum metallicity in the inner disc (see Figs.~\ref{MAPS79}, \ref{MAPS46} and \ref{MAPS13}) and similar changes in stellar age distributions, sketching a lasting imprint on the disc. In the case of $G_{\rm LMC}$, these effects peak for $\sim200$ Myr around the pericentre passages and last for up to 1 Gyr. If they are systematic and reliably modelled, they would provide a method to estimate the mass and orbit of a satellite galaxy that interacted with the main system, since the strength of the wave depends on these parameters. By analysing the surface brightness, metallicity, and stellar age distributions in simulations of isolated galaxies and comparing them with the observations, we can use the radial migration wave to infer independent constraints on satellite galaxy properties (e.g., its mass and orbit) and demonstrate the role of harassment as a satellite quenching mechanism \citep[e.g.,][]{rodriguezcardoso2025}. Additionally, we note longer-lasting impacts of tidally-induced migration in K9\textsubscript{i} ($Q=1.5$) with a $\sim4\%$ metallicity shift relative to K7, lowering it in the outer disc and enhancing it in the inner disc (see Fig. \ref{MAPS79}).

\section{Conclusions}
\label{sec:conclusions}
In this work, we leveraged the high-resolution simulation suite KRATOS \citepalias{jimenez2024kratos} to study radial migration under the influence of tidal interactions in LMC-like galaxies. These pure $N$-body simulations of the Clouds include a control group of isolated simulations of $G_{\rm LMC}$, an LMC-like galaxy and an interacting group where $G_{\rm LMC}$ interacts with the SMC-mass system $G_{\rm SMC}$ and the MW-mass system $G_{\rm SMC}$. We focus on a subsample including 6 of the 28 KRATOS simulations (see Table \ref{table:KRATOS}), chosen to be suited for a systematic comparison of the effects of the tidal interactions (see \ref{sec:kratos}). In order to quantify the impact of these interactions, we mapped the non-axisymmetric pattern strength, the net churning fluxes, and the impact of radial migration on the radial metallicity distribution in $G_{\rm LMC}$ in three interacting simulations and their three isolated counterparts. In this context, our key results can be summarised as follows:
\begin{enumerate}
      \item We present a description of a global radial migration wave pattern triggered by tidal interactions, along with comparative mapping of the effects of galactic encounters in LMC-like systems. Its amplitude can reach up to $\sim40\%$ of disc mass per Gyr at times close to the $G_{\rm LMC}$-$G_{\rm SMC}$ pericentre passages.
      \item These oscillating net churning fluxes come with a metallicity drop in the galactic disc, matching the pericentre passages of the satellite, and amounting to $\sim$3-5\% of the maximum value in the galaxy’s radial metallicity distribution prior to the interaction.
      \item During the interactions, all simulations behave similarly in the studied range of Toomre parameters ($1.0\leq Q \leq 1.5$), suggesting that strong tidal interactions tend to govern the radial migration in galactic discs.
      \item Radial migration tends to behave differently in the region below corotation—which is dominated by bar rotation—than above corotation resonance.
\end{enumerate}

Using both simulations and observations, future studies could better characterize tidally-induced radial migration waves, which drive inward migration before pericentre passages and outward migration afterward. They may lead to enhanced surface brightness in the inner disc and reduced brightness in the outer disc during pericentre passages, along with the metallicity drop we presented in the inner disc and similar changes in stellar age distributions. If these biases are systematic and reliably modelled, they could provide a method to estimate the mass and distance of a satellite galaxy, since the strength of the wave depends on these parameters. With sufficiently precise observations and accurate models of surface brightness, metallicity, and stellar age distributions for isolated galaxies, the measured biases could yield independent constraints on satellite galaxy properties.

Within KRATOS's high-resolution framework, we uncovered oscillations of radial migration and estimated their imprint on the radial metallicity distribution in LMC-like galaxies and anticipated the biases they could induce on observational properties as surface brightness and stellar age distribution. Yet, our analysis is limited to the migration of immutable stars, overlooking their rich environment, asymmetries, evolutions, births and deaths. As a result, future efforts must integrate these missing elements to reveal novel insights on the complexity of galactic dynamics, relying on combining simulations and observations. $Gaia$ DR4, expected December 2026, will improve the precision of its kinematic data for nearly two billion stars. In parallel, high-resolution hydrodynamical simulations may improve the theoretical modelling of stellar kinematics. The integration of these approaches will advance our understanding of the dynamical processes that govern disc galaxies.

\begin{acknowledgements}
      This work is based on the Master's thesis of DH (2025, Université de Toulouse) conducted at Lund University's Division of Astrophysics, under the supervision of OJA and SRF. The computer ``Brigit'' in the ``Centro de Datos de la Universidad Complutense de Madrid'' was used to re-run KRATOS simulations with an increased time cadency. OJA acknowledges funding from ``Swedish National Space Agency 2023-00154 David Hobbs The GaiaNIR Mission'' and ``Swedish National Space Agency 2023-00137 David Hobbs The Extended Gaia Mission''.
\end{acknowledgements}

\bibliographystyle{aa}
\bibliography{biblio.bib}

\begin{appendix}

\section{Additional maps}
\begin{figure*}[b]
    \centering
    \includegraphics[width=0.86\textwidth]{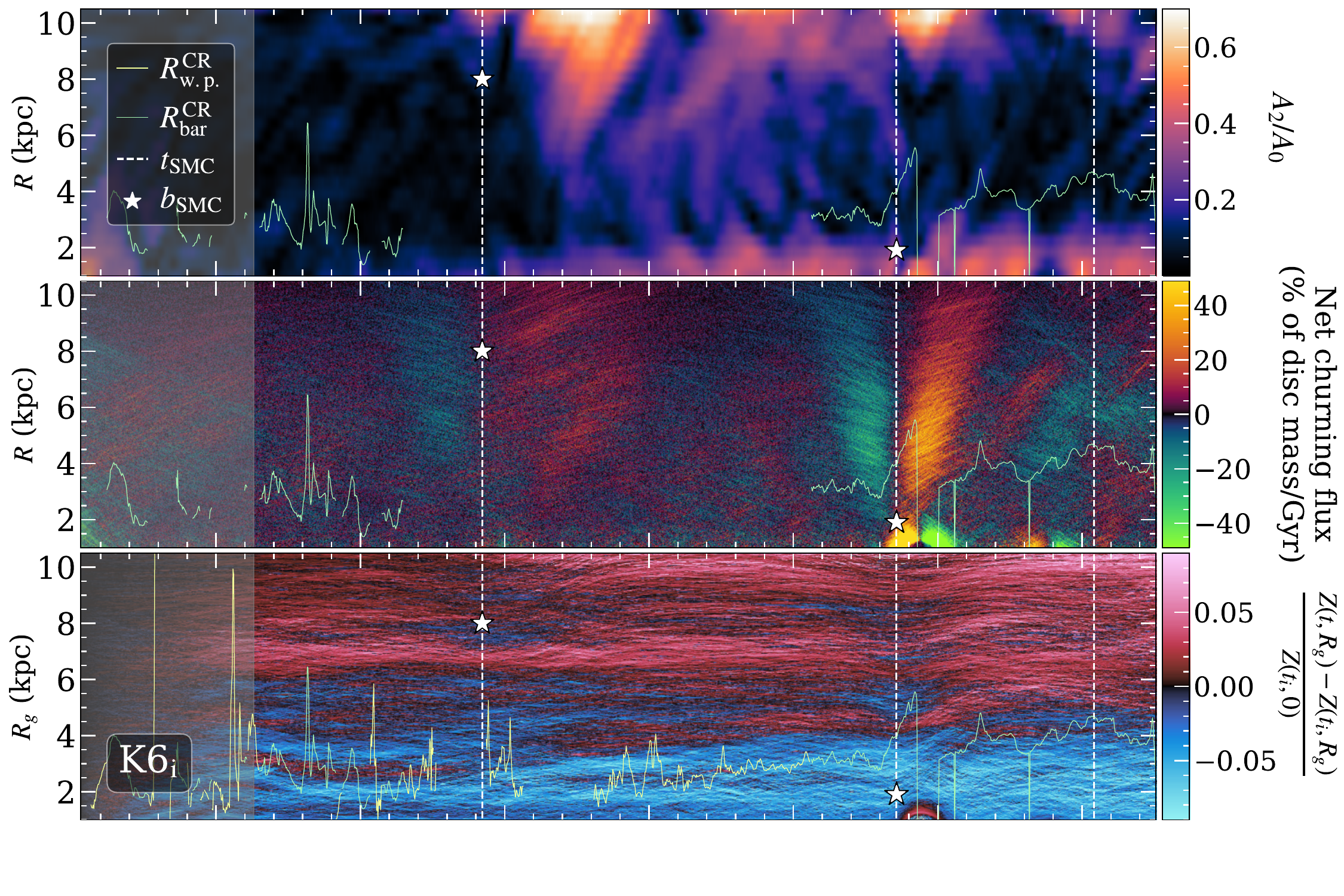}
    \includegraphics[width=0.86\textwidth]{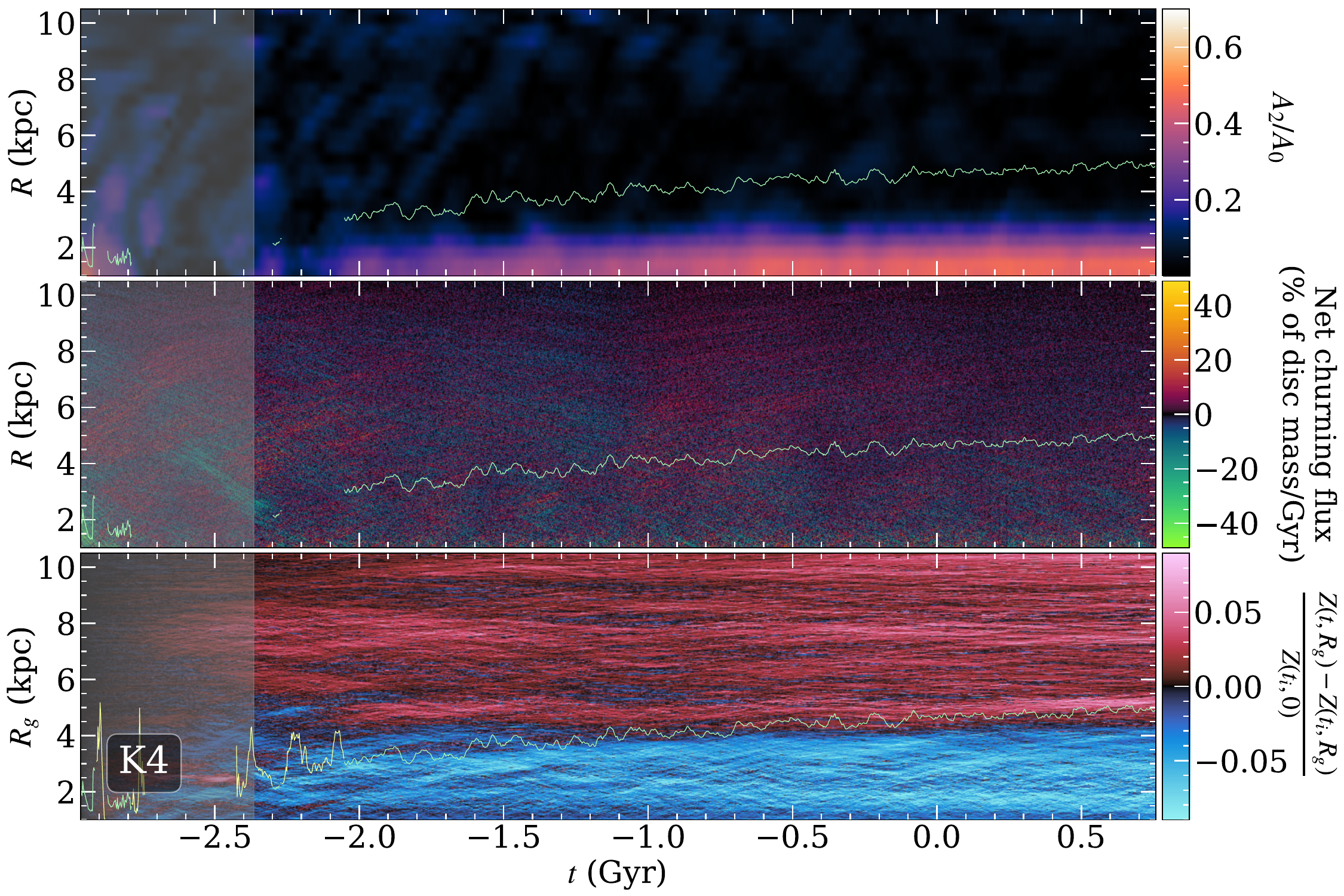}
    \caption{Same as Fig.~\ref{MAPS79} for K6\textsubscript{i} (top three panels, $Q=1$ interacting simulation) and K4 (bottom three panels, $Q=1$ isolated simulation).}
    \label{MAPS46}
\end{figure*}

\clearpage
\begin{figure*}
    \centering
    \includegraphics[width=0.98\textwidth]{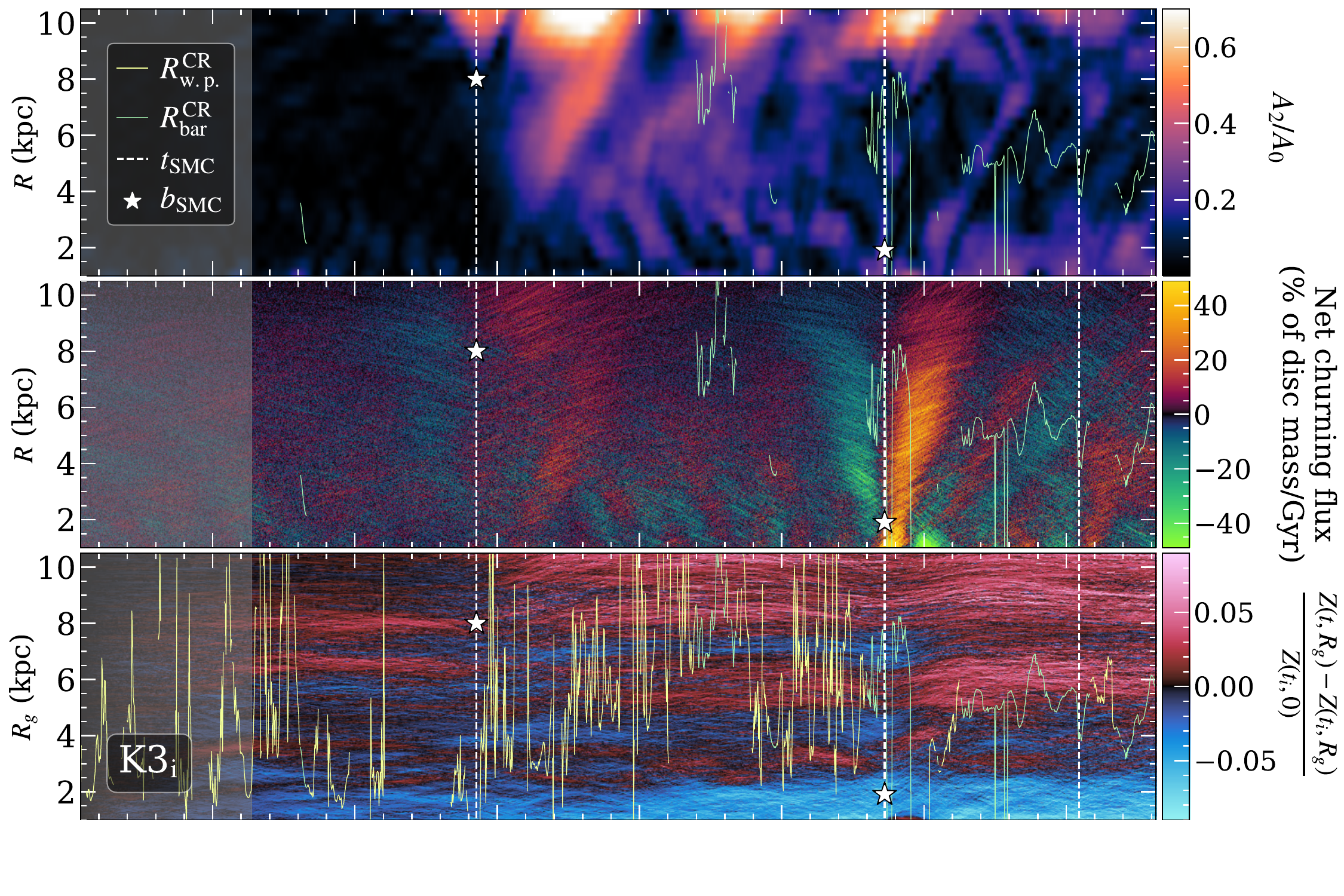}
    \includegraphics[width=0.98\textwidth]{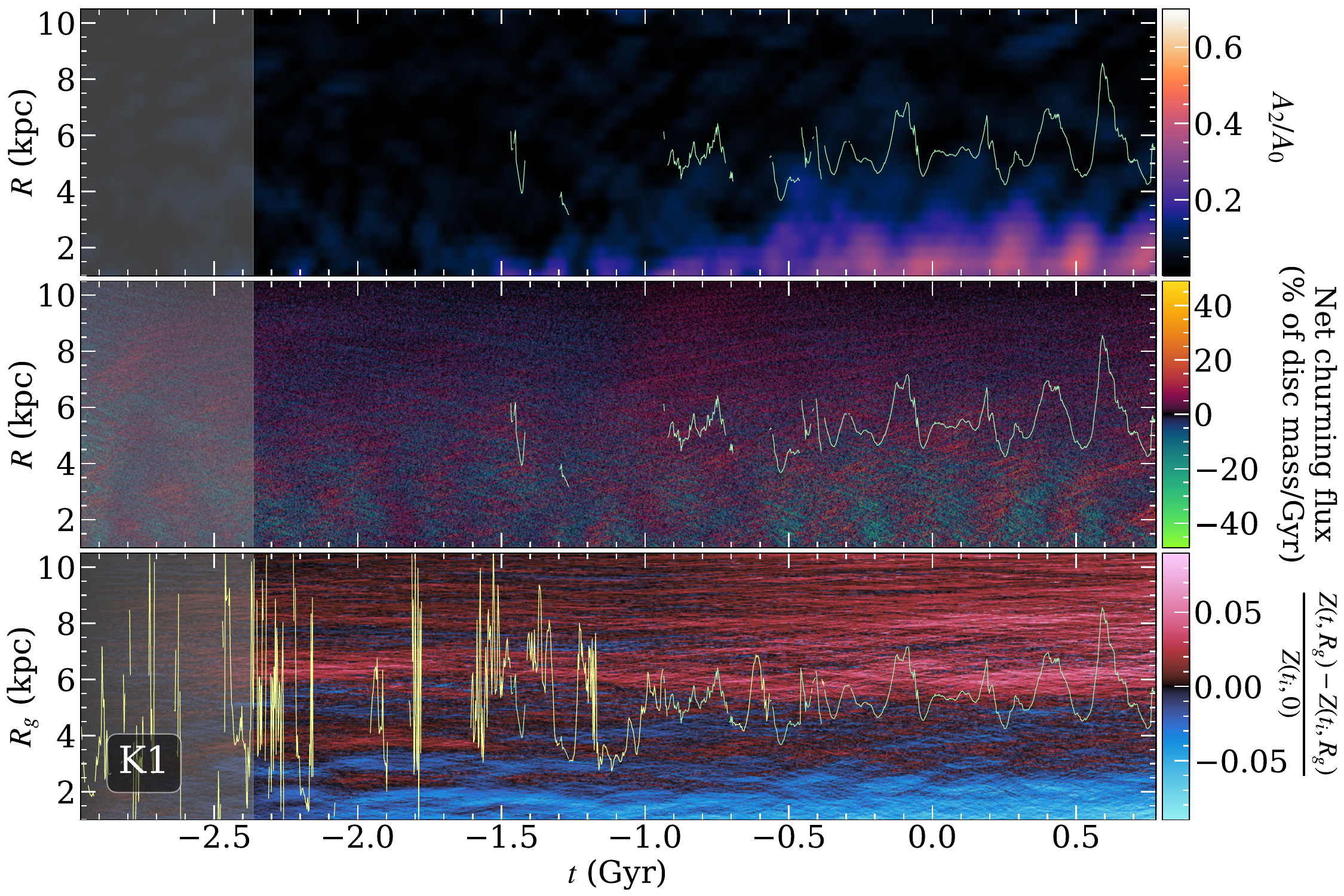}
    \caption{Same as Figs.~\ref{MAPS79} and \ref{MAPS46} for K3\textsubscript{i} (top three panels, $Q=1.2$ interacting simulation) and K1 (bottom three panels, $Q=1.2$ isolated simulation).}
    \label{MAPS13}
\end{figure*}

\end{appendix}

\end{document}